\documentclass[journal,draftcls,onecolumn,12pt,twoside]{IEEEtran}
\usepackage{graphicx}
\usepackage{amsmath, amsfonts, amssymb, amsthm}
\usepackage{enumerate}
\usepackage{multirow}
\usepackage{color}
\theoremstyle{definition} \newtheorem{theorem}{Theorem}
\theoremstyle{definition} \newtheorem{corollary}[theorem]{Corollary}
\theoremstyle{definition} 
\theoremstyle{definition} 
\theoremstyle{definition} \newtheorem{lemma}[theorem]{Lemma}
\theoremstyle{definition} \newtheorem{algorithm}{Algorithm}

\theoremstyle{definition} \newtheorem*{convention}{Conventions}
\theoremstyle{definition} 
\theoremstyle{definition} \newtheorem*{example}{Example}
\newenvironment{Bullet}
{\begin{list}{}%
             {\setlength\labelsep{0pt}%
              \setlength\itemindent{0pt}%
              \setlength\leftmargin{18pt}%
              \setlength\labelwidth{18pt}%
              \setlength\topsep{0pt}%
              \setlength\parsep{0pt}%
              \setlength\itemsep{0pt}%
              }
\item[$\thickspace(*)\thickspace$\hfill]}%
{\end{list}}

\begin{document}
%

\sloppy

\title{Multicast Network Coding and Field Sizes\thanks{A preliminary version of this manuscript has been submitted to ISIT 2014.}}
\author{\IEEEauthorblockA{Qifu~(Tyler)~Sun\textsuperscript{\dag},~Xunrui~Yin\textsuperscript{\ddag},~Zongpeng~Li\textsuperscript{\ddag},~and~Keping~Long\textsuperscript{\dag}}\\
{\normalsize \IEEEauthorblockA{\textsuperscript{\dag}Institute of Advanced Networking Technology and New Service, \\
 University of Science and Technology Beijing, Beijing, P. R. China \\
\textsuperscript{\ddag} Department of Computer Science, University of Calgary, Calgary, Canada
}}
}

\maketitle

\begin{abstract}
In an acyclic multicast network, it is well known that a linear network coding solution over GF($q$) exists when $q$ is sufficiently large. In particular, for each prime power $q$ no smaller than the number of receivers, a linear solution over GF($q$) can be efficiently constructed. In this work, we reveal that a linear solution over a given finite field does \emph{not} necessarily imply the existence of a linear solution over all larger finite fields. Specifically, we prove by construction that: (i) For every source dimension no smaller than 3, there is a multicast network linearly solvable over GF(7) but not over GF(8), and another multicast network linearly solvable over GF(16) but not over GF(17); (ii) There is a multicast network linearly solvable over GF(5) but not over such GF($q$) that $q > 5$ is a Mersenne prime plus 1, which can be extremely large; (iii) A multicast network linearly solvable over GF($q^{m_1}$) and over GF($q^{m_2}$) is \emph{not} necessarily linearly solvable over GF($q^{m_1+m_2}$); (iv) There exists a class of multicast networks with a set $T$ of receivers such that the minimum field size $q_{min}$ for a linear solution over GF($q_{min}$) is lower bounded by $\Theta(\sqrt{|T|})$, but not every larger field than GF($q_{min}$) suffices to yield a linear solution. The insight brought from this work is that not only the field size, but also the order of subgroups in the multiplicative group of a finite field affects the linear solvability of a multicast network.
\end{abstract}


\section{Introduction}
Consider a multicast network, which is a finite directed acyclic multigraph with a unique source node $s$ and a set $T$ of receivers. Every edge in the network represents a noiseless transmission channel of unit capacity. The source generates $\omega$ data symbols belonging to a fixed symbol alphabet and will transmit them to all receivers via the network. The maximum flow, which is equal to the number of edge-disjoint paths, from $s$ to every receiver is assumed to be no smaller than $\omega$. The network is said to be solvable if all receivers can recover all $\omega$ source symbols based on their respective received data symbols. When the network has only one receiver, it is solvable by network routing. When $|T| > 1$, the paradigm of network routing does not guarantee the network to be solvable. The seminal paper \cite{AhlswedeLiYeung00} introduced the concept of network coding (NC) and proved that the considered multicast network has a NC solution over some infinitely large symbol alphabet. It was further shown in \cite{LiYeungCai03} that \emph{linear} NC suffices to yield a solution when the symbol alphabet is algebraically modeled as a sufficiently large finite field, and every intermediate node transmits a linear combination of its received data symbols over the symbol field. Since then, there have been extensive studies on the field size requirement of a linear solution for a multicast network.

From an algebraic approach, reference \cite{KoetterMedard03} first showed that a multicast network has a linear solution over GF($q$) as long as the prime power $q$ is larger than $\omega$ times the number $|T|$ of receivers. The requirement of $q$ for the existence of a linear solution over GF($q$) is further relaxed by \cite{HoKargerMedardKoetter03} to be larger than $|T|$, and such a solution can be efficiently constructed by the algorithm proposed in \cite{HarveyKargerMurota05}. %
Meanwhile, the efficient algorithm in \cite{Jaggi05} is able to construct a linear solution over GF($q$) when $q$ is no smaller than $|T|$, and hence this condition on $q$ is slightly relaxed compared with the ones in \cite{HoKargerMedardKoetter03} and \cite{HarveyKargerMurota05}. This efficient algorithm requires one to initially identify, for each receiver in $T$, $\omega$ edge-disjoint paths starting from the source and ending at it. Denote by $\eta$ the maximum number of paths among the $\omega|T|$ paths that contain a common edge. The parameter $\eta$ is always no larger than $|T|$. By a more elaborate argument, the algorithm in \cite{Jaggi05} is refined in \cite{Yeung12} such that it can construct a linear solution over GF($q$) as long as $q$ is no smaller than $\eta$.

Denote by $q_{min}$ the minimum field size for the existence of a linear solution over GF($q_{min}$), and by $q^*_{max}$ the maximum field size for the non-existence of a linear solution over GF($q^*_{max}$) (if the network is linearly solvable over every finite field, then $q^*_{max}$ is not well defined and we set it to 1 as a convention.) The algorithm in \cite{Jaggi05} implies that $|T|$ is an upper bound on $q^*_{max}$. For the special case that the source dimension $\omega$ is equal to 2, the upper bound on $q^*_{max}$ is reduced to $O(\sqrt{|T|})$ in \cite{Fragouli06}. %
In both cases, the upper bounds additionally guarantee the existence of a linear solution over every GF($q$) with $q$ larger than the bounds. On the other hand, references \cite{Feder03} and \cite{Lehman04} independently constructed a class of multicast networks with $q_{min}$ lower bounded by $\Theta(\sqrt{|T|})$. On any network in this class, a linear solution exists over every GF($q$) with $q$ no smaller than $q_{min}$, that is, $q^*_{max} < q_{min}$. These results indicate that in many cases,

\vspace{1mm}
\begin{Bullet}
{\em A multicast network that is linearly solvable over a given finite field \emph{GF}($q$) is also linearly solvable over every larger finite field.}
\end{Bullet}
\vspace{1mm}

\begin{figure}[htbp]
\centering
\scalebox{0.68}
{\includegraphics{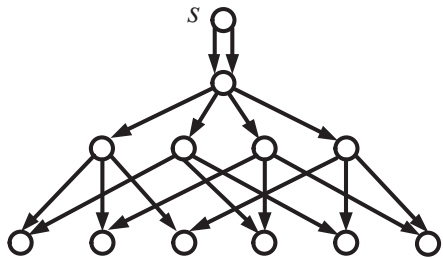}}
\caption{The $(4, 2)$-combination network has a unique source with $\omega = 2$ and 6 receivers at the bottom. It has a nonlinear solution over a ternary symbol alphabet but no solution over any symbol alphabet of size 6. It has a linear solution over every GF($q$) with $q \geq 3$.}
\label{Fig:4-2-Combination_Networks}
\end{figure}

To the best of our knowledge, there is no explicit proof or disproof of the above claim for a general multicast network and for the case $q < |T|$; furthermore, all known multicast networks studied in the network coding literature satisfy $q^*_{max} < q_{min}$. Although it has been shown in \cite{DFZeger04} that the $(4,2)$-combination (multicast) network depicted in Fig.~\ref{Fig:4-2-Combination_Networks} has a nonlinear solution over a ternary symbol alphabet but has neither a linear nor a nonlinear solution over any symbol alphabet of size 6, it does not shed light on disproving the claim $(*)$ because this combination network has a linear solution over every GF($q$) with $q \geq 3$ (See, for example, \cite{Sun_PIEEE11}.) Moreover, in the case $\omega = 2$, as revealed in \cite{Lehman04} and \cite{Fragouli06}, there exists a linear solution over GF($q$) if and only if there exists a $(q-1)$-vertex coloring in an appropriately defined associated graph. Since a $q$-vertex coloring in a graph always guarantees a $q'$-vertex coloring with $q' > q$ in the same graph, a multicast network with $\omega = 2$ is linearly solvable over every GF($q'$) with $q' \geq q_{min}$. This evidence seemed to add more support for the correctness of the claim $(*)$ for an arbitrary multicast network.

In the present paper, we shall show by constructive proofs that the claim $(*)$ is \emph{not} always true. In particular,
\begin{itemize}
\item we show that there is a multicast network that is linearly solvable over GF($q_{min}$) but not over GF($q^*_{max}$) for each of: (i) $q_{min} = 5$, $q^*_{max} = 8$; (ii) $q_{min} = 7$, $q^*_{max} = 8$; (iii) $q_{min} = 16$, $q^*_{max} = 17$;
\item we show that for any positive integer $d$ with less than 17,425,170 (base-10) digits, there is a multicast network with $q_{min} = 5$ whereas $q^*_{max} > d$. 
\item we construct a new class of multicast networks with $q_{min}$ lower bounded by $\Theta(\sqrt{|T|})$ and with an additional property that not every network in it has $q^*_{max}$ smaller than $q_{min}$. 
\end{itemize}

The insight of our results is that not only the field size but also the orders of the proper multiplicative subgroups in the symbol field affect the linear solvability over the finite field. As we shall see, if a finite field does not contain a large enough proper multiplicative subgroup, or the complement of a large multiplicative subgroup in the finite field is not large enough, it is possible to construct a multicast network that is not linearly solvable over this finite field but linearly solvable over a smaller finite field. In comparison, the characteristic of the symbol field does not appear as important in designing examples here as in the ones in \cite{DFZeger05} which show the non-existence of a linear solution for a general multi-source network, because in our exemplifying networks, both $q_{min}$ and $q_{max}^*$ can be of either odd and even characteristic. The classical solvable (non-multicast) network that is not linearly solvable, proposed in \cite{DFZeger05}, makes use of two types of subnetworks: one is linearly solvable only over a field with even characteristic whereas the other is linearly solvable only over a field with odd characteristic. Consequently, the proposed network as a whole is not linearly solvable over any field, even or odd. Our results bring about a new facet on the connection between the symbol field structures and network coding solvability problems.

The remainder of this paper is organized as follows. Section \ref{Sec:Fundamental_Results} presents the fundamental results that there exist multicast networks such that $q_{min} < q^*_{max}$. Section \ref{Sec:Gap} discusses how large the gap $q^*_{max} - q_{min}$ can be. %
To unify the justification of the linear solvability of different networks presented, Section \ref{Sec:General_Network} constructs a general multicast network and obtains an equivalent condition of its linear representability over a field GF($q$). Section \ref{Sec:Lower_bound} further constructs a new class of multicast networks with $q_{min}$ lower bounded by $\Theta(\sqrt{|T|})$ and with the additional property that $q_{min} < q^*_{max}$ for some networks in it. Section \ref{Sec:Summary} concludes the paper and discusses some interesting problems along this new research thread in network coding theory.

\section{Fundamental Results}\label{Sec:Fundamental_Results}
\begin{convention}
A (single-source) multicast network is a finite directed acyclic multigraph with a unique source node $s$ and a set $T$ of receivers. On a multicast network, for every node $v$, denote by $In(v)$ and $Out(v)$, respectively, the set of its incoming and outgoing edges. Every edge in a multicast network represents a noiseless transmission channel of unit capacity. The source generates $\omega$ data symbols belonging to a fixed symbol field and will transmit them to all receivers in $T$. Without loss of generality (WLOG), assume that $|Out(s)| = \omega$, which is referred to as the \emph{source dimension} (otherwise a new source can be created, connected to the old source with $\omega$ edges.) For an arbitrary set $N$ of non-source nodes, denote by $maxflow(N)$ the maximum number of edge-disjoint paths starting from $s$ and ending at nodes in $N$. Each receiver $t$ in $T$ has $maxflow(t) = \omega$.

A linear network code (LNC) over GF($q$) is an assignment of a \emph{coding coefficient} $k_{d,e} \in \mathrm{GF}(q)$ to every pair $(d, e)$ of edges such that $k_{d,e} = 0$ when $(d, e)$ is not an \emph{adjacent pair}, that is, when there is not a node $v$ such that $d \in In(v)$ and $e \in Out(v)$. The LNC uniquely determines a \emph{coding vector} $f_e$, which is an $\omega$-dim column vector, for each edge $e$ in the network such that:
\begin{itemize}
\item $\{f_e, e \in Out(s)\}$ forms the natural basis of $\mathrm{GF}(q)^\omega$.
\item $f_e = \sum_{d\in In(v)} k_{d,e}f_d$ when $e \in Out(v)$ for some $v \neq s$.
\end{itemize}
WLOG, we assume throughout this paper that
\begin{itemize}
\item all LNCs on a given multicast network have coding coefficients $k_{d,e} = 1$ for all those adjacent pairs $(d, e)$ where $d$ is the unique incoming edge to some node.
\end{itemize}
A multicast network is said to be \emph{linearly solvable} over GF($q$) if there is an LNC over GF($q$) such that for each receiver $t \in T$, the $\omega \times |In(t)|$ matrix $[f_e]_{e\in In(t)}$ over GF($q$) is full rank. Such an LNC is called a \emph{linear solution} over GF($q$) for the multicast network. Denote by $q_{min}$ the minimum field size for the existence of a linear solution over GF($q_{min}$), and by $q^*_{max}$ the maximum field size for the nonexistence of a linear solution over GF($q^*_{max}$). Specific to a finite field GF($q$), let GF($q$)$^\times$ represent the multiplicative group of nonzero elements in GF($q$). \hfill $\square$
\end{convention}

When $q \geq |T|$, the efficient algorithm in \cite{Jaggi05} can be adopted to construct a linear solution over GF($q$). In the case $\omega = 2$, it has been shown in \cite{Lehman04,Fragouli06} that every linear solution over GF($q$) can induce a $(q-1)$-vertex coloring in an appropriately associated graph of the multicast network and vice versa. Since every $(q-1)$-vertex coloring in a graph can also be regarded as a $(q'-1)$-vertex coloring for $q' > q$, it in turn induces a linear solution over GF($q'$) when $q'$ is a prime power. Thus, every linear solution over a finite field can induce a linear solution over a larger finite field. All these are tempting facts for one to conjecture that when $\omega > 2$ and $q < |T|$, a linear solution over a finite field might also imply the existence of a linear solution over a larger field. The central theme of the present paper is to \emph{refute} this conjecture in several aspects.

\begin{figure}[htbp]
\centering
\scalebox{0.5}
{\includegraphics{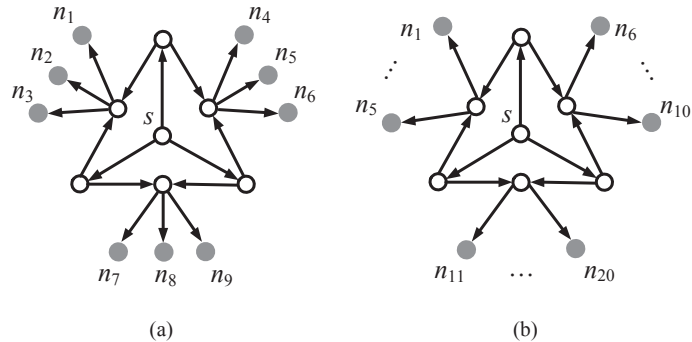}}
\caption{For both multicast networks, the source dimension $\omega$ is 3. There are totally 9 grey nodes in (a) and 20 grey nodes in (b). For every set $N$ of 3 grey nodes that has $maxflow(N) = 3$, there is a receiver connected from it, which is omitted in the depiction for simplicity. The network (a) is linearly solvable over every finite field GF($q$) with $q \geq q_{min} = 7$ except for $q = q^*_{max} = 8$. The network in (b) is linearly solvable over every finite field GF($q$) with $q \geq q_{min} = 16$ except for $q = q^*_{max} = 17$.}
\label{Fig:Rank_3_networks}
\end{figure}

\begin{theorem}\label{theorem:1}
A multicast network with $\omega \geq 3$ that is linearly solvable over a finite field is \emph{not} necessarily linearly solvable over all larger finite fields.

\begin{proof}
When $\omega = 3$, this theorem is a direct consequence of the lemma below. Assume $\omega > 3$. Expand the network $\mathcal{N}$ depicted in Fig.~\ref{Fig:Rank_3_networks}(a) to a new multicast network $\mathcal{N'}$ as follows. Create $\omega - 3$ new nodes each of which has an incoming edge emanating from the source and an outgoing edge entering every receiver. In $\mathcal{N'}$, every receiver has the maximum flow from the source equal to $\omega$. Consider an LNC over a given GF($q$). By the topology of $\mathcal{N'}$, the coding vector for every edge that is originally in $\mathcal{N}$ is a linear combination of the coding vectors for those edges in $Out(s)$ that are also in $\mathcal{N}$. Since the network $\mathcal{N}$ is linearly solvable over GF($q$) with every $q \geq 7$ except for $q = 8$ by the lemma below, so is the network $\mathcal{N'}$.
\end{proof}
\end{theorem}

\begin{lemma}\label{lemma:Rank_3_network_a}
Consider the multicast network depicted in Fig.~\ref{Fig:Rank_3_networks}(a). Denote by $e_i$, $1$ $\leq$ $i$ $\leq$ $9$ the unique incoming edge to node $n_i$. For every set $N$ of 3 grey nodes with $maxflow(N) = 3$, there is a receiver connected from it. This network is linearly solvable over every finite field GF($q$) with $q \geq q_{min} = 7$ except for $q = q^*_{max} = 8$.
\end{lemma}

A detailed technical proof for Lemma \ref{lemma:Rank_3_network_a} is given in Appendix \ref{Appendix_Proof_Rank_3_network_a}.

In the network in Fig.~\ref{Fig:Rank_3_networks}(a), observe that every receiver is connected with three nodes $n_i, n_j, n_k$ where either $\lceil i/3 \rceil$, $\lceil j/3 \rceil$, $\lceil k/3 \rceil$ are distinct (such as $\{n_1, n_4, n_7\}$) or two among $\lceil i/3 \rceil$, $\lceil j/3 \rceil$, $\lceil k/3 \rceil$ are same (such as $\{n_1, n_2, n_4\}$). One can then check an LNC over GF(7) with
\[
[f_{e_1}~\cdots~f_{e_9}]
= \left[
                         \begin{array}{ccccccccc}
                           1 & 1 & 1 & 0 & 0 & 0 & 1 & 1 & 1 \\
                           1 & 2 & 4 & 1 & 1 & 1 & 0 & 0 & 0 \\
                           0 & 0 & 0 & 1 & 2 & 4 & 1 & 2 & 4 \\
                         \end{array}
                       \right],
\]
where $e_j$ represents the unique incoming edge to node $n_j$, qualifies to be a linear solution for the network. Note that all nonzero entries in this matrix belong to a proper subgroup of order 3 in the multiplicative group $\mathrm{GF}(7)^\times$. The key reason that results in the network not linearly solvable over GF(8) is that there is not a proper subgroup of order at least 3 in the multiplicative group GF(8)$^\times$. %
In general, the network in Fig.~\ref{Fig:Rank_3_networks}(a) is linearly solvable over a finite field GF($q$) if and only if there exist $\alpha_i, \beta_i, \delta_i \in \mathrm{GF}(q)^\times, 1 \leq i \leq 3$ subject to
\begin{equation*}
\begin{aligned}
&\alpha_i \neq \alpha_j, \beta_i \neq \beta_j, \delta_i \neq \delta_j,~\forall~1 \leq i < j \leq 3, \\
&\{\delta_1, \delta_2, \delta_3\} \subseteq \mathrm{GF}(q)^\times\backslash \{-\alpha_i\beta_j: 1 \leq i,j \leq 3\}.
\end{aligned}
\end{equation*}
This is a direct consequence of Theorem \ref{thm:general_network_solvability} in Section \ref{Sec:General_Network} when a more general multicast network depicted in Fig.~\ref{Fig:General_Network} is to be introduced which subsumes the network in Fig.~\ref{Fig:Rank_3_networks}(a) as a special instance.

On a solvable multi-source network, it is well-known that linear network coding (with linearity in terms of a more general algebraic structure) is not sufficient to yield a solution. The classical example in \cite{DFZeger05} to show this consists of two subnetworks, one is only linearly solvable over a field with even characteristic, whereas the other is only linearly solvable over a field with odd characteristic. More generally, a procedure is introduced in \cite{DFZeger07} to construct a (matroidal) multi-source network based on a matroid such that the constructed network is linearly solvable over GF($q$) if and only if the matroid is representable over GF($q$). This connection is powerful for designing a number of non-linearly solvable networks from a variety of interesting matroid structures (See the Appendix in \cite{Oxley_Book} for example,) and the network in \cite{DFZeger05} is a instance under this construction. %
Subsequently, the role of characteristics of finite fields on the linear solvability of a general multi-source network is further revealed in \cite{DFZeger08}: for an arbitrary finite or co-finite set $S$ of prime numbers, a multi-source network can be constructed such that the network is linearly solvable over some field of characteristic $p$ for every prime number $p$ in $S$, whereas it is not linearly solvable over any finite field whose characteristic is not in $S$. %
However, the examples presented in this paper cannot be established by the procedures introduced in \cite{DFZeger07} and \cite{DFZeger08}, because they were not designed to construct (single-source) multicast networks. %
Moreover, as a result of Lemma \ref{lemma:Rank_3_network_a} and the next lemma, the role of the characteristic of a finite field in the examples designed in the present paper is not as important as in the example in \cite{DFZeger05}.

\begin{lemma}\label{lemma:Rank_3_network_b}
Consider the multicast network depicted in Fig.~\ref{Fig:Rank_3_networks}(b). There are in total 20 grey nodes. For every set $N$ of 3 grey nodes with $maxflow(N) = 3$, there is a receiver connected from it. This network is linearly solvable over every finite field GF($q$) with $q \geq q_{min} = 16$ except for $q = q^*_{max} = 17$.
\end{lemma}

A detailed proof of this lemma can be found in Appendix \ref{Appendix_Proof_Rank_3_network_b}. %
Similar to the network in Fig.~\ref{Fig:Rank_3_networks}(a), the network in Fig.~\ref{Fig:Rank_3_networks}(b) can also be regarded as a special instance of the general network to be introduced in Section \ref{Sec:General_Network}. Then, Theorem \ref{thm:general_network_solvability} to be developed in Section \ref{Sec:General_Network} implies that %
the network in Fig.~\ref{Fig:Rank_3_networks}(b) is linearly solvable over a finite field GF($q$) if and only if there is an assignment of $\{\alpha_i, \beta_i\}_{1\leq i \leq 5}$, $\{\delta_i\}_{1 \leq i \leq 10}$ from GF($q$)$^\times$ subject to
\begin{eqnarray*}
\alpha_i \neq \alpha_j,\beta_i \neq \beta_j, \delta_k \neq \delta_l, \nonumber \forall~1 \leq i < j \leq 5, 1 \leq k < l \leq 10,
\end{eqnarray*}
and
\begin{equation*}
\{\delta_1, \cdots, \delta_{10}\} \subseteq \mathrm{GF}(q)^\times\backslash \{-\alpha_i\beta_j: 1 \leq i,j \leq 5\}.
\end{equation*}
For example, when $q = 16$, we can set $\alpha_i = \beta_i = \xi^{3i}$ for all $1 \leq i \leq 5$, where $\xi$ is a primitive element in GF(16), and set $\delta_1, \cdots, \delta_{10}$ to be the 10 elements in $\mathrm{GF}(16)^\times\backslash\{\xi^{3i}: 1 \leq i \leq 5\}$. Such an assignment satisfies the above condition and hence the network is linearly solvable over GF(16). %
The insight for the network in Fig.~\ref{Fig:Rank_3_networks}(b) to be linearly solvable over GF(16) rather than GF(17) is that there is such a subgroup in the multiplicative group GF(16)$^\times$ but not in GF(17)$^\times$ that (i) the subgroup has order no smaller than 5; (ii) the complement of the subgroup in the multiplicative group contains at least 10 elements.

\begin{corollary}\label{Corollary:Characteristic}
Given a multicast network with $q_{min} < q^*_{max}$, both $q_{min}$ and $q^*_{max}$ can be of either even or odd characteristic.
\begin{proof}
This can be seen from Lemma \ref{lemma:Rank_3_network_a} together with Lemma \ref{lemma:Rank_3_network_b}.
\end{proof}
\end{corollary}

\section{Gap between $q_{min}$ and $q^*_{max}$}\label{Sec:Gap}
To the best of our knowledge, all known multicast networks studied in the network coding literature have the property that $q^*_{max} < q_{min}$.
The results in the previous section reveal that it is possible for $q_{min} < q^*_{max}$. However, both examples which illustrate this fact have the special property $q^*_{max} = q_{min} + 1$. A natural question next is how far away can $q^*_{max}$ be from $q_{min}$. The main result in this section is to show that for some multicast networks, the difference $q^*_{max} - q_{min}$ can be extremely large.

\begin{figure}[htbp]
\centering
\scalebox{0.58}
{\includegraphics{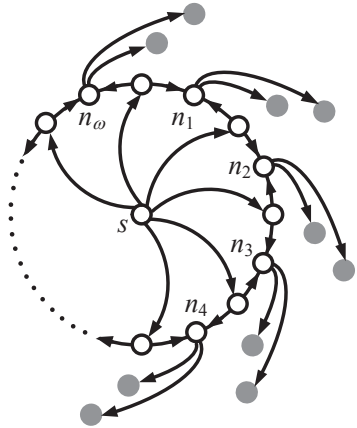}}
\caption{The \emph{Swirl network} has source dimension $\omega\geq 3$. Corresponding to every set $N$ of $\omega$ grey nodes that has $maxflow(N) = \omega$, there is a non-depicted receiver connected from it. For each node $n_i$ of in-degree 2, $1\leq i \leq \omega$, let $e_{i1}, e_{i2}$ denote the two outgoing edges from it.}
\label{Fig:Swirl_network}
\end{figure}

Consider the \emph{Swirl Network} with source dimension $\omega \geq 3$ depicted in Fig.~\ref{Fig:Swirl_network}. For every set $N$ of $\omega$ grey nodes with $maxflow(N) = \omega$, there is a non-depicted receiver connected from it. Corresponding to each node $n_i$ of in-degree 2, where $1\leq i \leq \omega$, let $e_{i1}, e_{i2}$ denote the two outgoing edges from it.
%

By viewing the Swirl network as a special instance of the general network to be introduced in the next section, the following necessary and sufficient condition for the Swirl Network to be linearly solvable over GF($q$) can be readily obtained. A detailed proof is given in Appendix \ref{Appendix_Proof_Swirl_network}.

\begin{lemma}\label{lemma:Swirl_Network}
The Swirl network is linearly solvable over a given GF($q$) if and only if there exist $\alpha_1, \cdots, \alpha_{\omega-1}, \delta_1, \delta_2 \in \mathrm{GF}(q)^\times$ subject to
\begin{eqnarray}
\label{eqn:Swirl_1}
&& \delta_1 \neq \delta_2~\mathrm{and}~\alpha_j \neq 1 ~\mathrm{for~all~} 1\leq j \leq \omega-1, \\
&~& \{\delta_1, \delta_2\} \subseteq \mathrm{GF}(q)^\times \backslash \{(-1)^\omega\gamma_1\gamma_2\cdots\gamma_{\omega-1}:  \nonumber \\
\label{eqn:Swirl_2}
&~& ~~~~~~~~~~~~ \gamma_i \in \{1, \alpha_j\}~\mathrm{for~all~} 1\leq j \leq \omega-1\}.
\end{eqnarray}
under which a linear solution over GF($q$) can be subsequently given by prescribing the coding vectors for $\{e_{i1}, e_{i2}\}_{1\leq i \leq \omega}$ as
\begin{equation}\label{eqn:Swirl_matrix}
[f_{e_{11}}~f_{e_{12}}~f_{e_{21}}~f_{e_{22}} \cdots f_{e_{\omega1}}~f_{e_{\omega2}}]
= \left[
  \begin{matrix}
    1 & 1 & 0 & 0 & \cdots & 0 & 0 & 1 & 1 \\
    1 & \alpha_1 & 1 & 1 & \ddots & 0 & 0 & 0 & 0 \\
    0 & 0 & 1 & \alpha_2 & \ddots & 0 & 0 & 0 & 0 \\
    \vdots & \vdots & \vdots & \vdots & \ddots & \vdots & \vdots & \vdots & \vdots \\
    0 & 0 & 0 & 0 & \cdots  & 1 & 1 & 0 & 0 \\
    0 & 0 & 0 & 0 & \cdots  & 1 & \alpha_{\omega-1} & \delta_1 & \delta_2 \\
  \end{matrix}
\right].
\end{equation}
\end{lemma}

\begin{example}
Consider the Swirl network with $\omega = 6$, and an LNC over GF(5) with the coding vectors for $\{d_1, e_1, \cdots d_6, e_6\}$ prescribed by
\[
[f_{d_1}~f_{e_1}\cdots f_{d_6}~f_{e_6}]
= \left[
  \begin{array}{cccccccccccc}
    1 & 1 & 0 & 0 & 0 & 0 & 0 & 0 & 0 & 0 & 1 & 1 \\
    1 & 4 & 1 & 1 & 0 & 0 & 0 & 0 & 0 & 0 & 0 & 0 \\
    0 & 0 & 1 & 4 & 1 & 1 & 0 & 0 & 0 & 0 & 0 & 0 \\
    0 & 0 & 0 & 0 & 1 & 4 & 1 & 1 & 0 & 0 & 0 & 0 \\
    0 & 0 & 0 & 0 & 0 & 0 & 1 & 4 & 1 & 1 & 0 & 0 \\
    0 & 0 & 0 & 0 & 0 & 0 & 0 & 0 & 1 & 4 & 2 & 3 \\
  \end{array}
\right].
\]
Apparently condition (\ref{eqn:Swirl_1}) is satisfied by this LNC. Since $\{1, 4\}$ is a subgroup of GF(5)$^\times$, it is closed under multiplication by elements in it. Moreover, $\{2, 3\} = \mathrm{GF}(5)^\times \backslash \{(-1)^6\cdot 1, (-1)^6\cdot 4\}$, and so condition (\ref{eqn:Swirl_2}) is also satisfied. Thus, this LNC over GF(5) qualifies as a linear solution. On the other hand, since GF(8)$^\times$ does not have a proper subgroup other than $\{1\}$, it is not difficult to check that for arbitrary $\alpha_1, \cdots, \alpha_5 \in \mathrm{GF}(8)^\times\backslash\{1\}$, the set $\{\gamma_1\gamma_2\cdots\gamma_{5}: \gamma_i \in \{1, \alpha_i\}~\mathrm{for~all~} 1\leq i \leq 5\}$ contains at least 6 elements. Thus, there are not enough distinct elements in GF(8)$^\times$ to assign for $\alpha_1, \cdots, \alpha_5$ and $\delta_1, \delta_2$ subject to conditions (\ref{eqn:Swirl_1}) and (\ref{eqn:Swirl_2}). Hence, the network is not linearly solvable over GF(8).
\end{example}

The argument in the example above can simply be generalized to derive the linear solvability of the Swirl network with general source dimension $\omega$ over a given GF($q$). First, assume that the order of the multiplicative group GF($q$)$^\times$ is not prime. This implies $q \geq 5$. Then there is a subgroup $G$ in GF($q$)$^\times$ with $|G| \geq 2$, and GF($q$)$^\times\backslash\{(-1)^\omega g: g \in G\}$ contains no less than 2 elements. Let $a$ be an element in $G$ not equal to 1 and $b, c$ be two distinct elements in GF($q$)$^\times\backslash\{(-1)^\omega g: g \in G\}$. We can assign $\alpha_i = a$ for all $1 \leq i \leq \omega - 1$ and $\delta_1 = b, \delta_2 = c$. Such an assignment obeys conditions (\ref{eqn:Swirl_1}) and (\ref{eqn:Swirl_2}). Hence, the network is linearly solvable over GF($q$).
Next, assume that the order of GF($q$)$^\times$ is a prime, in which case $q$ can always be written in the form of $2^p$ for some prime $p$. We shall show that for any $1 \leq a_1, a_2, \cdots, a_{n} \leq 2^p-2$, the set $S_n = \{b_1+\cdots+b_n \mod 2^p-1: b_i \in \{0, a_i\} \}$ contains at least $\min\{n+1, 2^p-1\}$ elements. This is obviously true when $n = 1$. Assume that it is true for $n = m$ and consider the case $n = m+1$. Note that $S_n = S_m \bigcup \{a_n+b \mod 2^p-1: b \in S_m\}$. If $|S_m| \geq 2^p-1$, then so is $|S_n|$. Assume that $2^p-1 > |S_m| \geq m+1$. Since $a_n \neq 0$ and $2^p-1$ is a prime, there is at least one element in $\{a_n+b \mod 2^p-1: b \in S_m\}$ that is not in $S_m$. Thus, $|S_n| \geq |S_m|+1 \geq n+1$. As a consequence, we conclude that for any assignment of $\alpha_1 = \xi^{a_1}, \cdots, \alpha_{\omega-1} = \xi^{a_{\omega-1}}$, where $\xi$ is a primitive element in GF($2^p$) and $1 \leq a_1, \cdots, a_{\omega-1} \leq 2^p-2$, the set $\{\gamma_1 \cdots \gamma_{\omega-1}: \gamma_i \in \{1, \xi^{a_i}\} \}$ contains at least $\omega$ elements. In order to further successfully assign $\delta_1, \delta_2$ subject to (\ref{eqn:Swirl_1}) and (\ref{eqn:Swirl_2}), the size of GF($2^p$) has to be at least $\omega + 3$. When $2^p \geq \omega+3$, we can set $\alpha_i = \xi$ for all $1 \leq i \leq \omega-1$ and $\delta_1 = (-1)^\omega\xi^\omega$, $\delta_2 = (-1)^\omega\xi^{\omega+1}$, where $\xi$ is a primitive element in GF($2^p$), so that conditions (\ref{eqn:Swirl_1}) and (\ref{eqn:Swirl_2}) obey.
We have proved the following theorem.

\begin{theorem}\label{theorem:swirl}
The Swirl network with $\omega \geq 3$ has $q_{min} = 5$ and $q^*_{max} = \max\{2^p: 2^p-1\mathrm{~is~prime}, 2^p \leq \omega+2\}$. Moreover, it is not linearly solvable over all those $2^p \leq q^*_{max}$ when $2^p - 1$ is prime.
\end{theorem}

Recall that a prime integer in the form $2^p - 1$ is known to be a \emph{Mersenne prime}. Then $q^*_{max}$ for the Swirl network with $\omega \geq 3$ is equal to one plus the largest Mersenne prime no larger than $\omega + 1$. While whether there are infinitely many Mersenne primes is still an open problem, the $48^{th}$ known Mersenne prime (which is also the largest known prime) found under the GIMPS project (See \cite{GIMPS}) has a length of 17,425,170 digits under base 10. Thus, when $\omega$ is sufficiently large, the difference $q^*_{max} - q_{min}$ for the Swirl network is so enormous as to having tens of millions of digits.

\begin{corollary}\label{corollary:swirl}
If there are infinitely many Mersenne primes, then there are infinitely many multicast networks with $q^*_{max} > q_{min}$, and moreover, the difference $q^*_{max} - q_{min}$ can tend to infinity.
\end{corollary}

Another interesting consequence of Theorem \ref{theorem:swirl} is that the Swirl network is linearly solvable over GF($2^4$), but not over a larger GF($2^p$) when $2^p - 1 \leq \omega+1$ is a prime. This unveils that the linear solvability over a finite field does not even guarantee the linear solvability over all larger finite fields of the \emph{same} characteristic. Moreover, it was pointed out in \cite{Fragouli11}, without demonstrating an explicit example or rigorous proof, that a multicast network linearly solvable over GF($q^{m_1}$) and over GF($q^{m_2}$) may not be linearly solvable over GF($q^{m_1 + m_2}$). The Swirl network is the first class of explicit networks that asserts this conjecture to be correct.

\begin{corollary}\label{corollary:swirl_characteristic}
There exists a multicast network such that it is linearly solvable over GF($q^{m_1}$) and over GF($q^{m_2}$) but \emph{not} over GF($q^{m_1+m_2}$).
\begin{proof}
Assume $\omega \geq 2^{13} = 2^4\cdot2^9$. Since $2^{13} - 1$ is a prime whereas $2^4-1, 2^9-1$ are not, the Swirl network is linearly solvable over GF($2^4$) and over GF($2^9$) but not over GF($2^{13}$).
\end{proof}
\end{corollary}

\noindent \textbf{Remark.}
The inspiration of designing the class of Swirl networks stems from a matroid structure referred to as the \emph{free swirl} (See \cite{Oxley_FreeSwirl} or Chapter 14 in \cite{Oxley_Book}). The matrix in (\ref{eqn:Swirl_matrix}) can be regarded as a (linear) representation of the free swirl of rank $\omega$ when conditions (\ref{eqn:Swirl_1}) and (\ref{eqn:Swirl_2}) hold. In a multicast network, the coding vectors of an LNC naturally induces a representable matroid on the edge set, and in the strongest sense, thus the induced representable matroid is referred to as a \emph{network matroid} \cite{Sun_ISIT08}, in which case the linear independence of coding vectors coincides with the independence structure of edge-disjoint paths \cite{Sun_CL13}. Similar to a network matroid, a \emph{gammoid} in matroid theory characterizes the independence of node-disjoint paths in a directed graph. It turns out that if the independence structure of a matroid can be reflected on a multicast network, then the matroid should be not only representable, but (isomorphic to) a gammoid as well. Being representable and a gammoid is the essential structure of the free swirl that guides the construction of multicast networks we look for. In comparison, those representable matroids, such as Fano and non-Fano matroids that have well-known applications to construct a multi-source network \cite{DFZeger07}, are not isomorphic to any gammoid, and hence they do not have corresponding multicast networks reflecting their independence structures.

\section{A General 5-Layer Multicast Network}\label{Sec:General_Network}
One may observe that the networks presented in the previous two sections with $q_{min} < q^*_{max}$ share a similar layered structure. In order to unify the derivation of their respective linear solvability, we next construct a general multicast network with parameters $(\omega, d_1, d_2)$ which subsumes the networks depicted in Fig.~\ref{Fig:Rank_3_networks} and Fig.~\ref{Fig:Swirl_network} as special cases, and then derive an equivalent condition for the linear solvability of this general network over a field GF($q$).

\begin{algorithm}
\label{Alg:General_Construction}
Given a 3-tuple of positive integers ($\omega, d_1, d_2$) as input parameters, the procedure below constructs a multicast network consisting of nodes on five layers, which are labeled 1-5 from upstream to downstream, and all edges are between adjacent layers.

\vspace{5pt}

\noindent \underline{\emph{Step 1}}. Create a source $s$, which forms the unique node at layer 1.

\vspace{3pt}

\noindent \underline{\emph{Step 2}}. Create $\omega$ layer-2 nodes, each of which is connected with $s$ by an edge. Sequentially label these nodes as $u_1, u_2, \cdots, u_\omega$.

\vspace{3pt}

\noindent \underline{\emph{Step 3}}. Create $\omega$ layer-3 nodes, labeled as $v_1, v_2, \cdots, v_\omega$. Each node $v_i$, $1 \leq i \leq \omega$ has 2 incoming edges, one leading from $u_i$ and the other from $u_{i-1}$, where $u_0$ will represent $u_\omega$.

\vspace{3pt}

\noindent \underline{\emph{Step 4}}. For each layer-3 node $v_i$, $1 \leq i \leq \omega-1$, create $d_1$ downstream layer-4 nodes $n_{i, 1}, n_{i, 2}, \cdots, n_{i, d_1}$, each of which has the unique incoming edge $e_{ij}$ leading from $v_{i}$. Corresponding to the layer-3 node $v_{\omega}$, create $d_2$ downstream layer-4 nodes $n_{\omega, 1}, n_{\omega, 2}, \cdots, n_{\omega, d_2}$, each of which has the unique incoming edge $e_{\omega j}$ leading from $v_{\omega}$.

\vspace{3pt}

\noindent \underline{\emph{Step 5}}. For every set $N$ of $\omega$ layer-4 nodes with $maxflow(N) = \omega$, create a layer-5 node connected from every node in $N$ by an edge. Set all layer-5 nodes to be receivers.
\hfill $\square$
\end{algorithm}

\begin{figure}
\centering
\scalebox{0.58}
{\includegraphics{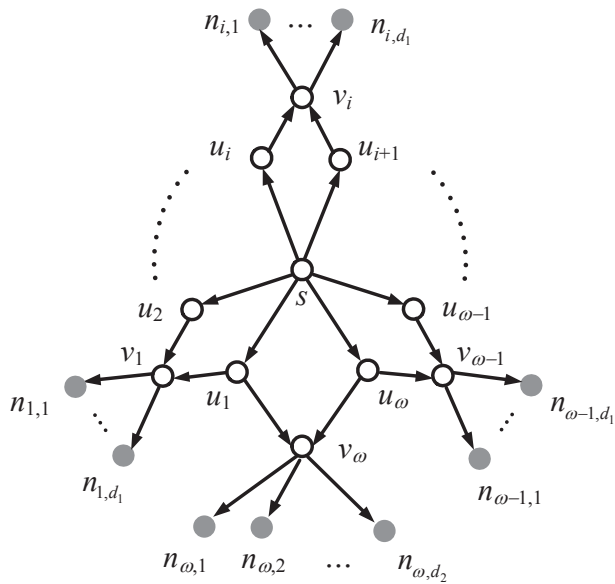}}
\caption{A general network constructed by Algorithm 1 based on parameters $(\omega, d_1, d_2)$ consists of nodes on 5 layers. The layer-1 just consists of the source node $s$, and all layer-4 nodes are depicted in grey. There is a non-depicted bottom-layer node connected from every set $N$ of $\omega$ layer-4 nodes with $maxflow(N) = \omega$. All bottom-layer nodes are receivers. The networks in Fig.~\ref{Fig:Rank_3_networks} and Fig.~\ref{Fig:Swirl_network} can be constructed by Algorithm 1 with appropriate settings of $(\omega, d_1, d_2)$.}
\label{Fig:General_Network}
\end{figure}

Fig.~\ref{Fig:General_Network} depicts the general multicast network constructed by Algorithm \ref{Alg:General_Construction} with parameters $(\omega, d_1, d_2)$. It can be easily checked that the networks in Fig.~\ref{Fig:Rank_3_networks}(a) and Fig.~\ref{Fig:Rank_3_networks}(b), as well as the Swirl network are all instances of the general network with respective parameters $(3, 3, 3)$, $(3, 5, 10)$, $(\omega, 2, 2)$.

\begin{theorem}
\label{thm:general_network_solvability}
Consider the general network constructed by Algorithm 1 with parameters $(\omega, d_1, d_2)$. The network is linearly solvable over a finite field GF($q$) if and only if there exist $\alpha_{ij}, \delta_{k} \in \mathrm{GF}(q)^\times$, $1 \leq i \leq \omega -1$, $1 \leq j \leq d_1$, $1 \leq k \leq d_2$ such that
\begin{align}
\label{eqn:general_network_linear_solvability}
\alpha_{ij} &\neq \alpha_{ij'}, \forall 1 \leq i \leq \omega -1, 1 \leq j < j' \leq d_1, \nonumber \\
\delta_{k} &\neq \delta_{k'}, \forall 1 \leq k < k' \leq d_2, \\
\delta_{1}, \cdots, \delta_{d_2} &\notin \big\{(-1)^\omega\gamma_1\gamma_2\cdots\gamma_{\omega-1}:\gamma_j \in \{\alpha_{j1}, \cdots, \alpha_{jd_1}\}~\forall 1 \leq j \leq \omega-1 \big\} \nonumber,
\end{align}
under which a linear solution over GF($q$) can be given by prescribing the coding vectors
\begin{align}
&~~\left[f_{e_{11}}, \cdots, f_{e_{1d_1}}, \cdots, f_{e_{(\omega-1)1}}, \cdots, f_{e_{(\omega-1)d_1}}, f_{e_{\omega 1}}, \cdots, f_{e_{\omega d_2}} \right] \nonumber \\
\label{eqn:general_network_matrix_reduced}
&= \left[
  \begin{array}{ccccccccccccc}
    1 & \cdots & 1 & 0 & \cdots & 0 &  & 0 & \cdots & 0 & 1 & \cdots  & 1 \\
    \alpha_{11} & \cdots & \alpha_{1d_1} & 1 & \cdots & 1 &  & \vdots & \ddots & \vdots & 0 & \cdots  & 0\\
    0 & \cdots & 0 & \alpha_{21} & \cdots & \alpha_{2d_1} &  & 0 & \cdots & 0 & 0 & \cdots & 0\\
    0 & \cdots & 0 & 0 & \cdots & 0 & \ddots & 0 & \cdots & 0 & \vdots & \ddots & \vdots \\
    \vdots & \ddots & \vdots & \vdots & \ddots & \vdots & & 1 & \cdots & 1 & 0 & \cdots & 0 \\
    0 & \cdots & 0 &  0 & \cdots & 0 &  & \alpha_{(\omega-1)1} & \cdots & \alpha_{(\omega-1)d_1} & \delta_{1} & \cdots & \delta_{d_2} \\
  \end{array}
\right].
\end{align}
\begin{proof}
On the network constructed by Algorithm \ref{Alg:General_Construction} with parameters $(\omega, d_1, d_2)$, consider an LNC with all coding coefficients being indeterminates. Assume the coding vector for the unique incoming edge to node $u_j$ is equal to the $j^{th}$ $\omega$-dim unit vector. Then, the juxtaposition of the coding vectors for edges $e_{ij}$, $1 \leq i \leq \omega$ can be represented as the $\omega \times ((\omega-1)d_1+d_2)$ matrix
\begin{align}
&~~\left[f_{e_{11}}, \cdots, f_{e_{1d_1}}, \cdots, f_{e_{(\omega-1)1}}, \cdots, f_{e_{(\omega-1)d_1}}, f_{e_{\omega 1}}, \cdots, f_{e_{\omega d_2}} \right] \nonumber \\
\label{eqn:matrix_coding_vectors}
&= \left[ \begin{array}{ccccccccccccc}
    \alpha_{11,1} & \cdots & \alpha_{1d_1,1} & 0 & \cdots & 0 &  & 0 & \cdots & 0 & \delta_{1,1} & \cdots  & \alpha_{d_2,1} \\
    \alpha_{11,2} & \cdots & \alpha_{1d_1,2} & \alpha_{21,1} & \cdots & \alpha_{2d_1,1} &  & \vdots & \ddots & \vdots & 0 & \cdots  & 0\\
    0 & \cdots & 0 & \alpha_{21,2} & \cdots & \alpha_{2d_1,2} &  & 0 & \cdots & 0 & 0 & \cdots & 0\\
    0 & \cdots & 0 & 0 & \cdots & 0 & \ddots & 0 & \cdots & 0 & \vdots & \ddots & \vdots \\
    \vdots & \ddots & \vdots & \vdots & \ddots & \vdots & & \alpha_{(\omega-1)1,1} & \cdots & \alpha_{(\omega-1)d_1,1} & 0 & \cdots & 0 \\
    0 & \cdots & 0 &  0 & \cdots & 0 &  & \alpha_{(\omega-1)1,2} & \cdots & \alpha_{(\omega-1)d_1,2} & \delta_{1,2} & \cdots & \delta_{d_2,2} \\
  \end{array}
\right]
\end{align}
where $\alpha_{ij,1}$, $\alpha_{ij,2}$ respectively represent the coding coefficient indeterminates for adjacent pairs $((u_i,v_i), e_{ij})$ and $((u_{i+1},v_i), e_{ij})$, $1 \leq i \leq \omega-1$, $1 \leq j \leq d_1$, and $\delta_{j,1}$, $\delta_{j,2}$ respectively represent the coding coefficient indeterminates for adjacent pairs $((u_\omega,v_\omega), e_{\omega j})$ and $((u_{1},v_\omega), e_{\omega j})$, $1 \leq j \leq d_2$. %
Denote by $K$ this juxtaposed matrix, and by $\Gamma$ the set of indeterminates $\alpha_{ij,m}, \delta_{k, m}$, where $1 \leq i \leq \omega -1$, $1 \leq j \leq d_1$, $1 \leq k \leq d_2$, and $m \in \{1, 2\}$. For every $\omega \times \omega$ submatrix in $K$, its determinant is a polynomial in indeterminates in $\Gamma$. Then, for every set $N$ of $\omega$ layer-4 nodes, the maximum flow from $s$ to $N$ is equal to $\omega$ if and only if the juxtaposition of coding vectors for incoming edges to nodes in $N$ has a nonzero determinant polynomial. Consequently, there is a linear solution over GF($q$) if and only if there is a \emph{matrix completion} for matrix $K$ over GF($q$), that is, an assignment of values in GF($q$) to indeterminates in $\Gamma$ such that for every $\omega \times \omega$ full-rank submatrix in $K$, the evaluation of its determinant polynomial is nonzero.

Assume that the network is linearly solvable over GF($q$). Then, there is a matrix completion for $K$ over GF($q$). Consider such a matrix completion that the indeterminates in $\Gamma$ are assigned to values in GF($q$).

Observe that the $\omega\times\omega$ matrix $[f_{e_{11}}, f_{e_{12}}, f_{e_{21}}, \cdots, f_{e_{(\omega-1)1}}] =
\left[\begin{smallmatrix}
\alpha_{11, 1} & \alpha_{12,1} & 0 & & 0 \\
\alpha_{11, 2} & \alpha_{12, 2} & \alpha_{21, 1} & & \cdots \\
0 & 0 & \alpha_{21, 2} & \cdots & 0 \\
\cdots & \cdots & 0 & & \alpha_{(\omega-1)1, 1} \\
0 & 0 & 0 &  & \alpha_{(\omega-1)1, 2}
\end{smallmatrix}\right]$ is full rank before matrix completion, and its determinant is equal to $(\alpha_{11, 1}\alpha_{12, 2} - \alpha_{11, 2}\alpha_{12, 1})\alpha_{21, 1}\cdots \alpha_{(\omega-1)1, 1}$. Hence, $\alpha_{11, 1}\alpha_{12, 2} \neq \alpha_{11, 2}\alpha_{12, 1}$ and $\alpha_{21, 1}, \cdots, \alpha_{(\omega-1)1, 1} \neq 0$. By an analogous argument to all those $\omega\times\omega$ full-rank submatrices of $K$ which are in the form $[f_{e_{ij}}, f_{e_{ij'}}, f_{e_{(i+1)j_1}}, \cdots, f_{e_{(i+\omega-2)j_{\omega-2}}}]$, where $e_{ij}$ represents $e_{(i-\omega)j}$ when $i > \omega$, it can be deduced that
\begin{itemize}
\item $\alpha_{ij, m}, \delta_{k, m} \neq 0$, $\forall 1 \leq i \leq \omega-1, 1\leq j \leq d_1, 1 \leq k \leq d_2, m\in \{1, 2\}$;
\item $\alpha_{ij, 1}\alpha_{ij', 2} \neq \alpha_{ij, 2}\alpha_{ij', 1}$, $\forall 1 \leq i \leq \omega-1, 1\leq j < j' \leq d_1$;
\item $\delta_{k, 1}\delta_{k', 2} \neq \delta_{k, 2}\delta_{k', 1}$, $\forall 1\leq k < k' \leq d_2$.
\end{itemize}

Thus, we can form another matrix $K'$ by the multiplication of $K$ and a diagonal square matrix with diagonal entries equal to $\{\alpha_{11, 1}^{-1}, \cdots, \alpha_{1d_1, 1}^{-1}, \cdots \alpha_{(\omega-1)1, 1}^{-1}, \cdots, \alpha_{(\omega-1)d_1, 1}^{-1}, \delta_{1, 1}^{-1}, \cdots, \delta_{d_2, 1}^{-1}\}$. Any $\omega$ columns in $K$ are linearly independent if and only if the $\omega$ columns in $K'$ of same indices are linearly independent. %
Now, set $\alpha_{ij} = \alpha_{ij,1}^{-1}\alpha_{ij,2}$, where $1 \leq i \leq \omega-1$, $1 \leq j \leq d_1$, and set $\delta_{k} = \delta_{k,1}^{-1}\delta_{k,2}$, where $1 \leq k \leq d_2$. Then, $K'$ is in the same form as expressed in (\ref{eqn:general_network_matrix_reduced}), and
\begin{itemize}
\item $\alpha_{ij}, \delta_{k} \neq 0$, $\forall 1 \leq i \leq \omega-1, 1\leq j \leq d_1, 1 \leq k \leq d_2$;
\item $\alpha_{ij} \neq \alpha_{ij'}$, $\delta_{k} \neq \delta_{k'}$ $\forall 1 \leq i \leq \omega-1, 1\leq j < j' \leq d_1, 1 \leq k < k' \leq d_2$.
\end{itemize}

Next, observe that the $\omega\times\omega$ submatrix $[f_{e_{11}}, f_{e_{21}}, \cdots, f_{e_{\omega 1}}] =
\left[\begin{smallmatrix}
\alpha_{11, 1} &  0 & & 0 & \delta_{1, 1}\\
\alpha_{11, 2} & \alpha_{21, 1} & & \cdots & 0 \\
0 & \alpha_{21, 2} & \cdots & 0 & \cdots \\
\cdots & 0 & & \alpha_{(\omega-1)1, 1} & 0 \\
0 & 0 &  & \alpha_{(\omega-1)1, 2} & \delta_{1, 2}
\end{smallmatrix}\right]$ of $K$ is of full rank before matrix completion. Hence, the $\omega\times \omega$ submatrix in $K'$ with the same column indices $\left[\begin{smallmatrix}
1 &  0 & & 0 & 1\\
\alpha_{11} & 1 & & \cdots & 0 \\
0 & \alpha_{21} & \cdots & 0 & \cdots \\
\cdots & 0 & & 1 & 0 \\
0 & 0 &  & \alpha_{(\omega-1)1} & \delta_{1}
\end{smallmatrix}\right]$ is of full rank. Equivalently, its determinant $\delta_{1} + (-1)^\omega\alpha_{11}\cdots\alpha_{(\omega-1)1}$ is nonzero. By the same argument to the full-rank $\omega \times \omega$ submatrices $[f_{e_{1j_1}}, f_{e_{2j_2}}, \cdots, f_{e_{\omega j_\omega}}]$, $1 \leq j_1, \cdots, j_{\omega-1} \leq d_1$ and $1 \leq j_\omega \leq d_2$, it can be deduced that
\[
\delta_{k} \neq (-1)^\omega \alpha_{1j_1}\cdots\alpha_{(\omega-1)j_{\omega-1}},~~\forall 1 \leq j_1, \cdots, j_{\omega-1} \leq d_1, 1 \leq k \leq d_2.
\]
Thus, the values $\alpha_{ij}, \delta_{k}$ induced from the considered matrix completion for $K$ over GF($q$) satisfies condition (\ref{eqn:general_network_linear_solvability}). The proof of the necessity part is complete.

To prove the sufficiency part, assume that there exist $\alpha_{ij}, \delta_{k} \in \mathrm{GF}(q)^\times$ subject to (\ref{eqn:general_network_linear_solvability}). Then, it can be carefully checked that the assignment
\[
\alpha_{ij, 1} = 1, \delta_{k, 1} = 1,\alpha_{ij, 2} = \alpha_{ij}, \delta_{k, 2} = \delta_k~\forall 1\leq i \leq \omega-1, 1\leq j \leq d_1, 1 \leq k \leq d_2
\]
qualifies as a matrix completion for the matrix $K$ depicted in (\ref{eqn:matrix_coding_vectors}) over GF($q$), and hence the matrix in (\ref{eqn:general_network_matrix_reduced}) is the juxtaposition of coding vectors from a linear solution over GF($q$).
\end{proof}
\end{theorem}

\begin{corollary}
The network depicted in Fig.~\ref{Fig:General_Network} with parameters $(\omega, d_1, d_2)$ is linearly solvable over GF($q$) if GF($q$)$^\times$ has a proper subgroup $G$ such that $|G| \geq d_1$ and $|\mathrm{GF}(q)^\times\backslash G| \geq d_2$.
\begin{proof}
Let $G$ be a subgroup in GF($q$)$^\times$ such that $|G| \geq d_1$ and $|\mathrm{GF}(q)^\times\backslash G| \geq d_2$. For each $1 \leq i \leq \omega-1$, assign $\alpha_{i1}, \alpha_{i2}, \cdots, \alpha_{id_1}$ to be arbitrary $d_1$ elements in $G$, %
and $\delta_1, \delta_2, \cdots, \delta_{d_2}$ to be arbitrary $d_2$ elements in $\mathrm{GF}(q)^\times\backslash G$. Under such assignments, $\big\{\gamma_1\gamma_2\cdots\gamma_{\omega-1}:\gamma_j \in \{\alpha_{j1}, \cdots, \alpha_{jd_1}\}~\forall 1 \leq j \leq \omega-1 \big\}$ is always contained in $G$ and thus condition (\ref{eqn:general_network_linear_solvability}) is satisfied and the general network is linearly solvable over GF($q$).
\end{proof}
\end{corollary}

\section{A New Class of Multicast Networks with $q_{min}$ lower bounded by $\Theta(\sqrt{|T|})$}\label{Sec:Lower_bound}
In the literature, it is well known that $|T|$ is an upper bound on $q_{min}$ for multicast networks. To the best of our knowledge, the highest lower bound on $q_{min}$ for multicast networks is of the order $\Theta(\sqrt{|T|})$. References \cite{Feder03} and \cite{Lehman04} independently constructed a class of multicast networks whose $q_{min}$ is lower bounded by $\Theta(\sqrt{|T|})$. The two classes of networks are essentially the same, and they are based on the class of $(n, 2)$-combination networks with $\omega = 2$ (The (4, 2)-combination network is depicted in Fig.~\ref{Fig:4-2-Combination_Networks}.) Moreover, they share the same property that $q^*_{max} < q_{min}$. Inspired by the results in Section \ref{Sec:Fundamental_Results}, we shall establish a new class of multicast networks with $q_{min}$ also lower bounded by $\Theta(\sqrt{|T|})$ whereas \emph{not} every network in it has $q^*_{max}$ smaller than $q_{min}$.

Let us first revisit the network depicted in Fig.~\ref{Fig:Rank_3_networks}(a). As proved in Lemma \ref{lemma:Rank_3_network_a}, it is linearly solvable over GF($q_{min}$) where $q_{min} = 7$ but not over GF(8). The network is assumed to have a receiver connected from every set $N$ of 3 grey nodes with $maxflow(N) = 3$. Actually, appropriate deletion of some receivers (e.g. the ones respectively connected from $\{n_2, n_3, n_4\}$ and $\{n_2, n_3, n_5\}$) from the network will not affect the linearly solvability over any particular finite field. Stemming from this observation, we shall next construct a new class of multicast networks with $q_{min}$ lower bounded by $\Theta(\sqrt{|T|})$ by appropriately increasing grey nodes and redefining receivers in the network in Fig.~\ref{Fig:Rank_3_networks}(a).

\begin{figure}
\centering
\scalebox{0.58}
{\includegraphics{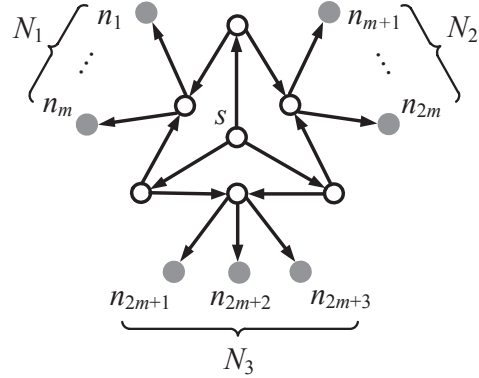}}
\caption{In the multicast network with source dimension 3, there are $2m+3$ grey nodes, which are classified into three sets $N_1, N_2, N_3$. Every receiver, which is not depicted for simplicity, is connected from some particular set of three grey nodes and there are a total of $4m^2+m+7$ receivers.}
\label{Fig:Network_LowerBound}
\end{figure}

Consider the network with $\omega = 3$ depicted in Fig.~\ref{Fig:Network_LowerBound}. It has a total of $2m+3$ grey nodes, which are classified into three sets as $N_1 = \{n_1, n_2, \cdots, n_m\}$, $N_2 = \{n_{m+1}, n_{m+2}, \cdots, n_{2m}\}$, and $N_3 = \{n_{2m+1}, n_{2m+2}, n_{2m+3}\}$. The set of receivers in this network is prescribed as follows:
\begin{itemize}
\item Type-I: Corresponding to every (unordered) pair $\{n_i, n_j\}$ of grey nodes in $N_1$, there is a receiver connected from $\{n_i, n_j, n_{m+k}\}$, where $k = 1$ when $i = 1$ and $k = j$ otherwise;
\item Type-II: Corresponding to every (unordered) pair $\{n_{m+i}, n_{m+j}\}$ of grey nodes in $N_2$, there is a type-II receiver connected from $\{n_k, n_{m+i}, n_{m+j}\}$, where $k = 1$ when $i = 1$ and $k = j$ otherwise;
\item Type-III: There is a respective receiver connected from $\{n_i, n_{2m+1},n_{2m+2}\}$ for every $n_i \in N_1$, from $\{n_{m+i}, n_{2m+1},n_{2m+3}\}$ for every $n_{m+i} \in N_2$, and from $\{n_1, n_{2m+2}, n_{2m+3}\}$. There is also a respective receiver connected from $\{n_1, n_2, n_{2m+i}\}$ and from $\{n_{m+1},n_{m+2}, n_{2m+i}\}$ for every $n_{2m+i} \in N_3$;
\item Type-IV: There is a receiver connected from $\{n_i, n_{m+j}, n_{2m+k}\}$ for every possible $n_i \in N_1, n_{m+j} \in N_2, n_{2m+k} \in N_3$.
\end{itemize}
The total number $|T|$ of receivers in the network is then equal to $2\left(\begin{matrix} m \\ 2 \end{matrix}\right) + 2m + 7 + 3m^2 = 4m^2 + m + 7$. Denote by $e_i$ the incoming edge to node $n_i$ for all $1 \leq i \leq 2m+3$.

The network depicted in Fig.~\ref{Fig:Network_LowerBound} can be viewed as an instance of the general network constructed by Algorithm 1 with parameters $(3, m, m)$, with some particularly chosen receivers deleted. By noticing that such careful deletion of receivers will not affect the linear solvability of the network, we can directly obtain the next lemma as a corollary of Theorem \ref{thm:general_network_solvability}.

\begin{lemma}\label{lemma:lower_bound}
The network depicted in Fig.~\ref{Fig:Network_LowerBound} is linearly solvable over a given GF($q$) if and only if there is an LNC over GF($q$) with coding vectors for $\{e_i\}_{1\leq i \leq 2m+3}$ prescribed by
\begin{equation}\label{eqn:Matrix_for_lower_bound}
\begin{aligned}
&~~~~[f_{e_1} \cdots f_{e_m}~f_{e_{m+1}}\cdots f_{e_{2m}}~f_{e_{2m+1}}\cdots f_{e_{2m+3}}] \\
&= \left[
  \begin{matrix}
  1 & \cdots & 1 & 0 & \cdots & 0 & 1 & 1 & 1 \\
  \alpha_1 & \cdots & \alpha_{m} & 1 & \cdots & 1 & 0 & 0 & 0 \\
  0 & \cdots & 0 & \beta_1 & \cdots & \beta_m & \delta_1 & \delta_2 & \delta_3
  \end{matrix}
\right],
\end{aligned}
\end{equation}
where
\begin{equation}\label{eqn:lower_bound_1}
\begin{aligned}
&\alpha_i, \beta_i \in \mathrm{GF}(q)^\times~~\forall 1 \leq i \leq m \\
&\alpha_i \neq \alpha_j, \beta_i \neq \beta_j~~\forall 1 \leq i < j \leq m,
\end{aligned}
\end{equation}
and
\begin{equation}\label{eqn:lower_bound_2}
\begin{aligned}
&\delta_1, \delta_2, \delta_3~\mathrm{are~distinct~elements~in} \\
&~~~~~~~~~~~\mathrm{GF}(q)^\times \backslash \{-\alpha_i\beta_j: 1 \leq i, j \leq m\}.
\end{aligned}
\end{equation}
\end{lemma}

In this lemma, prescribed by condition (\ref{eqn:lower_bound_1}), the set $\{-\alpha_i\beta_j: 1 \leq i, j \leq m\}$ contains at least $m$ elements. Hence, condition (\ref{eqn:lower_bound_2}) obeys only if $q \geq m+4$. Consider the special case that $m \geq 3$ and $2m+1$ is equal to a Mersenne prime. Since there is a subgroup $G$ of order $m$ in the multiplicative group GF($2m+1$)$^\times$, and $|\mathrm{GF}(2m+1)^\times\backslash G| = m \geq 3$, we are able to assign $\{\alpha_i, \beta_i\}_{1\leq i \leq m}$ and $\{\delta_j\}_{1 \leq j \leq 3}$ subject to (\ref{eqn:lower_bound_1}) and (\ref{eqn:lower_bound_2}) by setting $\{\alpha_i\}_{1\leq i \leq m} = \{\beta_i\}_{1\leq i \leq m} = G$ and $\delta_1, \delta_2, \delta_3$ to be arbitrary distinct elements in $\mathrm{GF}(2m+1)^\times\backslash \{-g: g\in G \}$. Therefore, the network is linearly solvable over GF($2m+1$). We next show that the network is not linearly solvable over GF($2m+2$). Recall the Cauchy-Davenport theorem (See \cite{Cauchy-Davenport} for example) which asserts that for any two nonempty subsets $A$ and $B$ in the additive group $\mathbb{Z}_p$, where $p$ is a prime, the cardinality of $A + B = \{a+b: a\in A, b\in B\}$ contains at least $\min\{|A|+|B|-1, p\}$ elements. Since the multiplicative group GF($2m+2$)$^\times$ is isomorphic to the additive group $\mathbb{Z}_{2m+1}$, where $2m+1$ is prime, there are at least $2m-1$ elements in $\{-\alpha_i\beta_j: 1 \leq i, j \leq m\}$ as a direct consequence of the Cauchy-Davenport theorem. Hence, there are only at most 2 elements in $\mathrm{GF}(2m+2)^\times\backslash \{-\alpha_i\beta_j: 1 \leq i, j \leq m\}$, so that condition (\ref{eqn:lower_bound_2}) cannot be satisfied. The above results are summarized as follows.

\begin{theorem}\label{theorem:lower_bound}
For the network depicted in Fig.~\ref{Fig:Network_LowerBound}, $q_{min} \geq m+4$, and hence $q_{min}$ is lower bounded by $\Theta(\sqrt{|T|})$. When $m \geq 3$ and $2m+1$ is equal to a Mersenne prime, $q_{min} \leq 2m+1$ whereas the network is \emph{not} linearly solvable over GF($2m+2$).
\end{theorem}

\section{Summary and Discussions}\label{Sec:Summary}
On an acyclic multicast network, if there is a linear solution over GF($q$), could we claim that there is a linear solution over every GF($q'$) with $q' \geq q$? It would be tempting to answer it positively because by the result in \cite{Jaggi05}, the claim is correct when $q$ is no smaller than the number of receivers and moreover, as a consequence of the result in \cite{Fragouli06}, the positive answer is affirmed for the special case that the source dimension of the network is equal to 2. In the present paper, however, we show the negative answer for general cases by constructing several classes of multicast networks with different emphasis. These networks are the first ones discovered in the network coding literature with the property that $q^*_{max}$, the maximum field size for the nonexistence of a linear solution over GF($q^*_{max}$), is larger than $q_{min}$, the minimum field size for the existence of a linear solution over GF($q_{min}$).

The insight of various exemplifying networks established in the present paper is that not only the field size of GF($q$), but also the \emph{order of the proper multiplicative subgroup} of GF($q$)$^\times$ affects the networks' linear solvability over GF($q$).

The results in this paper bring about a new thread on the fundamental study of linear network coding, specific to the case of multicast networks, as discussed below:
\begin{itemize}
\item Besides the field size and multiplicative subgroup orders, it is not clear whether there are some other undiscovered inherent structures in a finite field that affects the linear solvability of a multicast network. It deserves further investigation.
\item All multicast networks presented in this paper that are linearly solvable over a field GF($q$) but not over a larger field GF($q'$) share a common property that for some values $d_1, d_2$, there is a proper subgroup $G$ in GF($q$) subject to $|G| \geq d_1$ and $|\mathrm{GF}(q)^\times\backslash G| \geq d_2$, but there does not exist a proper subgroup $G'$ in GF($q'$) subject to $|G'| \geq d_1$ and $|\mathrm{GF}(q')^\times \backslash G| \geq d_2$. Thus, when $\mathrm{GF}(q)^\times$ does not contain any proper subgroup other than $\{1\}$, \emph{i.e.}, $q - 1$ is a prime, we can simply set $G = \{1\} \subset \mathrm{GF}(q)^\times$ and $G' = \{1\} \subset \mathrm{GF}(q')^\times$, such that for any values $d_1, d_2$ subject to $|G|\geq d_1$ and $|\mathrm{GF}(q)^\times \backslash G| \geq d_2$, the conditions $|G'|\geq d_1$ and $|\mathrm{GF}(q')^\times \backslash G'| \geq d_2$ hold as well. This leads us to an ambitious conjecture that if a multicast network is linearly solvable over GF($q$) where $q-1$ is a prime, then it is linearly solvable over all finite fields of sizes larger than $q$.
\item As special cases of the previous conjecture, it is of particular interest to study: i) whether a multicast network linearly solvable over GF(2) (or over GF(3)) is linearly solvable over all larger finite fields; ii) whether a multicast network linearly solvable over GF(2) and over GF(3) is linearly solvable over all finite fields. These two cases are related to other interesting conjectures in the network coding literature: It was conjectured and partially proved in \cite{Xiahou} that every planar multicast network is linearly solvable over GF(3), and certain special planar multicast networks (including relay-coface networks and terminal-coface networks) are always linearly solvable over both GF(2) and GF(3). The conjectures presented here will further suggest that these special multicast networks may be linearly solvable over all finite fields.
\end{itemize}

Stemming from the above discussions, we end this paper by proposing a number of open problems, all of which, except for the first, we conjecture to have positive answers:
\begin{itemize}
\item For a multicast network, what is the smallest prime power $q$ larger than $q^*_{max}$ (such that the network is linearly solvable over all GF($q'$) with $q' \geq q$)?
\item Can the gap $q^*_{max} - q_{min} > 0$ tend to infinity?
\item Are there infinitely many prime power pairs $(q, q')$ with $q < q'$ such that each $(q, q')$ corresponds to $(q_{min}, q^*_{max})$ of some multicast network?
\item If a multicast network is linearly solvable over such a GF($q$) that GF($q$)$^\times$ does not contain any proper multiplicative subgroup other than $\{1\}$, is it linearly solvable over all larger finite fields than GF($q$)?
\item If a multicast network is linearly solvable over GF(2) (or over GF(3)), is it linearly solvable over all larger finite fields?
\item If a multicast network is linearly solvable over both GF(2) and GF(3), is it linearly solvable over all finite fields?
\end{itemize}


\appendix
\subsection{Proof of Lemma \ref{lemma:Rank_3_network_a}}\label{Appendix_Proof_Rank_3_network_a}
Observe that the network in Fig.~\ref{Fig:Rank_3_networks}(a) can be constructed by Algorithm 1 introduced in Section \ref{Sec:General_Network} with parameters $(3, 3, 3)$. Then Theorem \ref{thm:general_network_solvability} implies that the network in Fig.~\ref{Fig:Rank_3_networks}(a) is linearly solvable over GF($q$) if and only if there exist $\alpha_{i}, \beta_{i}, \delta_{i} \in \mathrm{GF}(q)^\times$, $1 \leq i \leq 3$ such that
\begin{eqnarray}
\label{eqn:GF_7_GF_8_1}
&&\alpha_i \neq \alpha_j, \beta_i \neq \beta_j, \delta_i \neq \delta_j,~\forall~1 \leq i < j \leq 3, \\
&& \{\delta_1, \delta_2, \delta_3\} \subseteq \mathrm{GF}(q)^\times\backslash \{-\alpha_i\beta_j: 1 \leq i,j \leq 3\}.
\label{eqn:GF_7_GF_8_2}
\end{eqnarray}

Under constraint (\ref{eqn:GF_7_GF_8_1}), the cardinality of $\{-\alpha_i\beta_j: 1 \leq i,j \leq 3\}$ is no smaller than 3 and no larger than 9. Thus,
\begin{itemize}
\item when $q < 7$, there does not exist any assignment of $\{\alpha_i, \beta_i, \delta_i\}_{1 \leq i \leq 3}$ from GF($q$)$^\times$ satisfying (\ref{eqn:GF_7_GF_8_2}), and hence the network is not linearly solvable over GF($q$);
\item when $q \geq 13$, there are always ways to assign $\{\alpha_i, \beta_i, \delta_i\}_{1 \leq i \leq 3}$ from GF($q$)$^\times$ subject to (\ref{eqn:GF_7_GF_8_1}) and (\ref{eqn:GF_7_GF_8_2}), and hence the network is linearly solvable over GF($q$).
\end{itemize}
It remains to consider the case $q$ = 7, 8, 9, or  11.

When $q$ = 7, 9, or 11, there is a proper subgroup $G$ of order at least 3 in the multiplicative group $\mathrm{GF}(q)^\times$.
We can then assign arbitrary three distinct values in $G$ to $\alpha_1$, $\alpha_2$, $\alpha_3$ and to $\beta_1$, $\beta_2$, $\beta_3$. In this way, the cardinality of $\{-\alpha_i\beta_j: 1 \leq i,j \leq 3\}$ is upper bounded by $|G|$ and thus there are at least three values remained in $\mathrm{GF}(q)^\times\backslash \{-\alpha_i\beta_j: 1 \leq i,j \leq 3\}$, which $\delta_1$, $\delta_2$, $\delta_3$ can be assigned to.

In the last case $q$ = 8, since there is no proper subgroup in GF$(8)^\times$, the  method depicted in the previous paragraph does not work any more and an exhaustive search will verify that there does not exist any assignment of $\{\alpha_i, \beta_i, \delta_i\}_{1 \leq i \leq 3}$ from GF($q$)$^\times$ satisfying (\ref{eqn:GF_7_GF_8_1}) and (\ref{eqn:GF_7_GF_8_2}).

We can now affirm that the network depicted in Fig.~\ref{Fig:Rank_3_networks}(a) is linearly solvable over every GF($q$) with $q \geq q_{min} = 7$ except for $q = 8$.

\subsection{Proof of Lemma \ref{lemma:Rank_3_network_b}} \label{Appendix_Proof_Rank_3_network_b}
Observer that the network in Fig.~\ref{Fig:Rank_3_networks}(b) can be constructed by Algorithm 1 introduced in Section \ref{Sec:General_Network} with parameters $(3, 5, 10)$. Then Theorem \ref{thm:general_network_solvability} implies that the network in Fig.~\ref{Fig:Rank_3_networks}(b) is linearly solvable over GF($q$) if and only if there is an assignment of $\{\alpha_i, \beta_i\}_{1\leq i \leq 5}$, $\{\delta_i\}_{1 \leq i \leq 10}$ from GF($q$)$^\times$ subject to
\begin{eqnarray}
\label{eqn:GF_16_GF_17_1}
&& \alpha_i \neq \alpha_j,\beta_i \neq \beta_j, \delta_k \neq \delta_l, ~~~\forall~1 \leq i < j \leq 5, 1 \leq k < l \\
\label{eqn:GF_16_GF_17_2}
&&\{\delta_1, \cdots, \delta_{10}\} \subseteq \mathrm{GF}(q)^\times\backslash \{-\alpha_i\beta_j: 1 \leq i,j \leq 5\}.
\end{eqnarray}
Under condition (\ref{eqn:GF_16_GF_17_1}), the cardinality of $\mathrm{GF}(q)^\times\backslash \{-\alpha_i\beta_j: 1 \leq i,j \leq 5\}$ is lower bounded by 5 and upper bounded by 25. Thus,
\begin{itemize}
\item when $q < 16$, there does not exist any assignment of $\{\alpha_i, \beta_i\}_{1\leq i \leq 5}$, $\{\delta_i\}_{1 \leq i \leq 10}$  from GF($q$)$^\times$ satisfying (\ref{eqn:GF_16_GF_17_1}), and hence the network is not linearly solvable over GF($q$);
\item when $q > 36$, there are always ways to assign $\{\alpha_i, \beta_i\}_{1\leq i \leq 5}$, $\{\delta_i\}_{1 \leq i \leq 10}$ subject to (\ref{eqn:GF_16_GF_17_1}) and (\ref{eqn:GF_16_GF_17_2}), and hence the network is linearly solvable GF($q$).
\end{itemize}

It remains to consider the case that $16 \leq q < 36$. Note that when $q$ is not equal to 17 or 32, there is also a proper subgroup $G$ in the multiplicative group GF($q$)$^\times$ such that $G$ has order no smaller than 5 and the cardinality of GF($q$)$^\times\backslash G$ is at least 10. Then, we can respectively assign any 5 distinct elements in $G$ to $\{\alpha_i\}_{1\leq i \leq 5}$ and to $\{\beta_i\}_{1\leq i \leq 5}$, and assign any 10 distinct elements in GF($q$)$^\times\backslash \{-g: g \in G\}$ to $\{\delta_i\}_{1\leq i \leq 10}$. Such an assignment obeys conditions (\ref{eqn:GF_16_GF_17_1}) and (\ref{eqn:GF_16_GF_17_2}).

When $q = 32$, denote by $\xi$ be a primitive element in GF(32)$^\times$. Assign $\alpha_i = \beta_i = \xi^{2i}$ for all $1 \leq i \leq 5$, and $\delta_j = \xi^{2j-1}$ for all $1 \leq j \leq 10$. It is easy to check that such an assignment obeys conditions (\ref{eqn:GF_16_GF_17_1}) and (\ref{eqn:GF_16_GF_17_2}).

When $q = 17$, in order to make $\mathrm{GF}(17)^\times\backslash \{-\alpha_i\beta_j: 1 \leq i,j \leq 5\}$ contains at least 10 elements, $\{\alpha_i\}_{1\leq i \leq 5}$ and $\{\beta_i\}_{1\leq i \leq 5}$ should be such assigned that the cardinality of $\{\alpha_i\beta_j: 1 \leq i,j \leq 5\}$ is no larger than 6. To minimize this cardinality, as many as $\alpha_i$ and $\beta_j$ should be assigned to a same proper subgroup in GF(17)$^\times$. An exhaustive search will then verify that it is infeasible to assign $\{\alpha_i\}_{1\leq i \leq 5}$ and $\{\beta_i\}_{1\leq i \leq 5}$ from $\mathrm{GF}(17)^\times$ so that $\{\alpha_i\beta_j: 1 \leq i,j \leq 5\}$ contains no more than 6 elements.

We can now assert that the network depicted in Fig.~\ref{Fig:Rank_3_networks}(b) is linearly solvable over every GF($q$) with $q \geq q_{min} = 16$ except for $q = 17$.

\subsection{Proof of Lemma \ref{lemma:Swirl_Network}}\label{Appendix_Proof_Swirl_network}
Observe that the Swirl network can be constructed by Algorithm 1 in Section \ref{Sec:General_Network} with parameters $(\omega, 2, 2)$. Then according to Theorem \ref{thm:general_network_solvability}, the Swirl network is linearly solvable over GF($q$) if and only if there exist $\alpha_j, \beta_j, \delta_1, \delta_2 \in \mathrm{GF}(q)^\times$, $1\leq j \leq \omega-1$ subject to
\begin{eqnarray}
\label{eqn:Swirl_proof_1}
&&\delta_1 \neq \delta_2, \alpha_j \neq \beta_j ~~~\forall 1 \leq j \leq \omega-1, \\
\label{eqn:Swirl_proof_2}
&&\{\delta_1, \delta_2\} \subseteq \mathrm{GF}(q)^\times\backslash\{(-1)^\omega\gamma_1\cdots\gamma_{\omega-1}: \gamma_j \in \{\alpha_j, \beta_j\}\},
\end{eqnarray}
Given arbitrary $\alpha_j, \beta_j, \delta_1, \delta_2 \in \mathrm{GF}(q)^\times$, $1\leq j \leq \omega-1$, define  $\alpha_j' = \beta_j^{-1}\alpha_j \in \mathrm{GF}(q)^\times~~\forall 1 \leq j \leq \omega -1$ and $\delta_1' = \delta_1\prod_{i=1}^{\omega-1}\beta_i^{-1} \mathrm{GF}(q)^\times$,  $\delta_2' = \delta_2\prod_{i=1}^{\omega-1}\beta_i^{-1} \mathrm{GF}(q)^\times$. It is straightforward to check that condition (\ref{eqn:Swirl_proof_1}) holds if and only if
\[
\delta_1' \neq \delta_2', \alpha_j' \neq 1 ~~~\forall 1\leq j \leq \omega-1.
\]
Moreover, because
\begin{equation*}
\begin{aligned}
&\{\delta_1', \delta_2'\} = \{\delta_1\prod_{i=1}^{\omega-1}\beta_i^{-1}, \delta_2\prod_{i=1}^{\omega-1}\beta_i^{-1}\} \\ &\subseteq \mathrm{GF}(q)^\times\backslash\{(-1)^\omega\beta_1^{-1}\gamma_1\cdots\beta_{\omega-1}^{-1}\gamma_{\omega-1}: \gamma_j \in \{\alpha_j, \beta_j\}\} \\
& = \mathrm{GF}(q)^\times\backslash\{(-1)^\omega\gamma_1\cdots\gamma_{\omega-1}: \gamma_j \in \{1, \alpha_j'\}\},
\end{aligned}
\end{equation*}
condition (\ref{eqn:Swirl_proof_2}) holds if and only if
\[
\{\delta_1', \delta_2'\} \subseteq \mathrm{GF}(q)^\times \backslash \{(-1)^\omega\gamma_1\gamma_2\cdots\gamma_{\omega-1}: \gamma_j \in \{1, \alpha_j'\}\}.
\]
We can now conclude that the Swirl network is linearly solvable over GF($q$) if and only if there exist $\alpha_1, \cdots, \alpha_{\omega-1}, \delta_1, \delta_2 \in \mathrm{GF}(q)^\times$ satisfying conditions (\ref{eqn:Swirl_1}) and (\ref{eqn:Swirl_2}).

\section*{Acknowledgment}
The authors would like to thank Sidharth Jaggi for helping identify and motivate the problem studied in this work. The authors would also appreciate the helpful suggestions by the associate editor as well as anonymous reviewers to help improve the presentation of the paper.

\end{document}